\documentclass[aps,pra,twocolumn,superscriptaddress,showpacs,draft]{revtex4}
\input{epsf}

\newcommand{\rem}[1]{}

\newcommand{\ps}{\ps}

\newcommand{\Pop}{{\hat{\rm P}}}

\begin{document}

\title{Chaotic ratchet dynamics with cold atoms in a pair of 
pulsed optical lattices}

\author{Gabriel G. Carlo}
\affiliation{Center for Nonlinear and Complex Systems, Universit\`a degli 
Studi dell'Insubria, Via Valleggio 11, 22100 Como, Italy}
\affiliation{Departamento de F\'\i sica, Comisi\'on Nacional de Energ\'\i a 
At\'omica, Avenida del Libertador 8250, 1429 Buenos Aires, Argentina}

\author{Giuliano Benenti} 
\affiliation{Center for Nonlinear and Complex Systems, Universit\`a degli 
Studi dell'Insubria, Via Valleggio 11, 22100 Como, Italy}
\affiliation{Istituto Nazionale di Fisica Nucleare, Sezione di Milano
and CNISM}

\author{Giulio Casati} 
\affiliation{Center for Nonlinear and Complex Systems, Universit\`a degli 
Studi dell'Insubria, Via Valleggio 11, 22100 Como, Italy}
\affiliation{Istituto Nazionale di Fisica Nucleare, Sezione di Milano
and CNISM}
\affiliation{Department of Physics, National University of
Singapore, Singapore 117542, Republic of Singapore}

\author{Sandro Wimberger}
\author{Oliver Morsch}
\author{Riccardo Mannella}
\author{Ennio Arimondo}
\affiliation{CNR-INFM and Dipartimento di Fisica ``Enrico Fermi'',
Universit\`a degli Studi di Pisa, Largo Pontecorvo 3, 56127 Pisa, Italy}

\date{\today}

\pacs{05.45.Mt, 05.40.Jc, 05.60.-k, 32.80.Pj}
%05.45.Mt Quantum chaos; semiclassical methods
%05.40.Jc Brownian motion
%05.60.-k Transport processes
%32.80.Pj Optical cooling of atoms; trapping

\begin{abstract} 
We present a very simple model for realizing directed transport 
with cold atoms in a pair of periodically flashed optical lattices. 
The origin of this ratchet effect is 
explained and its robustness demonstrated under 
imperfections typical of cold atom experiments. 
We conclude that our model offers a clear-cut way to implement 
directed transport in an atom optical experiment. 
\end{abstract}

\maketitle

\section{Introduction}

The atom optics realization of the paradigmatic kicked rotor 
(KR) \cite{CCFI79}
presents the possibility to study experimentally unique quantum 
mechanical aspects of a fundamental, classically nonlinear system. 
Dynamical Localization is perhaps the most celebrated 
quantum phenomenon observed in the quantum KR \cite{CCFI79,localization}, 
but many other interesting features of the KR have been studied
theoretically and experimentally \cite{KRexperiment,KRcontrol}. Very 
recently, novel applications of modified KR models have been
designed which allow for a controlled, directed motion of
particles in momentum space \cite{cas,tanja}.

The atom-optics kicked rotor (AOKR) is realized by subjecting 
cold atoms  \cite{KRexperiment,KRcontrol} or a Bose condensate 
\cite{wilson,wilson1,pisa,nist} to a far detuned 
standing wave with spatial period $\pi/k_L$ 
($k_L$ being the wave number of the kicking laser) and pulsed 
with period $\tau$. 
The AOKR is described, in dimensionless units, 
by the Hamiltonian \cite{Graham1992}
\begin{equation}
{\cal H }(t) =\frac{p^2}{2} + k\cos(x)\sum_{n=0}^{\infty}
 \delta (t - n T)\;,
\label{eq:ham}
\end{equation}
where $p$ is the atomic momentum in units of $2\hbar k_L$ 
(i.e. of two--photon recoil momenta),
$x$ is the atomic position divided by $2k_L$, $t$ is time and $n$ is an
integer which counts the kicks. 
Experimentally, $\delta$--kicks are approximated by pulses of width $\tau_p$ 
which are approximately rectangular in shape. 
We also define an effective Planck's constant $\hbar_{\rm eff}=T = 8E_R\tau /\hbar$, 
where $E_R=(\hbar k_L)^2/2M$ is the recoil energy
(acquired by an atom after emission of a photon with wave number $k_L$). 
The dimensionless parameter $k \approx V_0\tau_p/\hbar$ 
is the kicking strength of the system (with $V_0$ the height of the
optical lattice creating the kicking potential). 

In this paper, we propose a ratchet which could be realized
experimentally by adding to the standard AOKR dynamics defined by
(\ref{eq:ham}) a second kicking potential (applied in a synchronized
way with respect to the first one). 
The application of a second kicking potential to the atom has
some analogy with the double AOKR investigated in \cite{jones04}, because
in both cases a sequence of two kicks is applied to the atoms. In the present
investigation a spatial shift of the second kick potential is also included.
We show that this is sufficient to produce the ratchet effect.
Moreover, we consider the effects of a particle escape 
mechanism similar to evaporative cooling \cite{Ket}.
%with the double AOKR.
%process
%substituting for dissipation.
%Dissipation in the AOKR was already introduced in \cite{cas}, but the present
%work introduces a new approach 
%which is somewhat 
%closer to an experimental
%realization with cold atoms, and combines a 
%mechanism similar to evaporative
%cooling \cite{Ket} with the double AOKR.
%To simulate the dissipative ratchet effect, we could use a zero temperature
%bath, as done in \cite{cas}. Much simpler to implement is, however, an
%effective opening of the system with absorbing boundary conditions.
More precisely, we study an open system with absorbing boundary conditions.
If $\psi(p)$ is the wave function in momentum space, absorbing boundary
conditions are implemented by the prescription $\psi(p)\equiv 0$ if
$p \leq -p_{c}$ or $p \geq p_{c}$. Such absorbing boundary conditions
could be realized experimentally using, e.g.,
velocity selective Raman transitions, which change the
internal states of the atoms, and hence let them escape from the states of
interest \cite{raman}, or by other state selective methods \cite{paris}. Such
a scenario of loosing the faster atoms with momenta exceeding $p_{c}$, is
analogous to evaporative cooling of cold atoms \cite{Ket}. The time-scale of the
applied absorption mechanism should be of the order
of the kicking period $T$ to allow for a steady loss of atoms during
the system's evolution.

We point out that, as shown below, in our model the ratchet phenomenon 
is also present in the Hamiltonian limit without escape of particles. 
On the other hand, it is interesting to investigate the particle escape 
mechanism because it models the evaporative cooling process natural 
in cold atoms experiments. Moreover, its introduction is 
relevant in order to analyze the stability of our proposed ratchet mechanism
after that atoms excited to higher and higher velocities by  
chaotic diffusion are eventually lost.
Finally, particle escape may allow the unprecedented experimental 
observation of a quantum phase space distribution located on an 
underlying classical fractal set.

In state-of-the-art atom optics experiments, 
control over the kicking strength $k$ (or, equivalently over the laser power 
delivered to the atoms) is achieved with a precision of a 
few percent \cite{KRcontrol}. Kicking strengths in the range 
$1\ldots 7$ correspond to standing wave amplitudes of about 
$80 \ldots 600 \; E_R$ for rubidium atoms (and assuming a rectangular pulse
shape with a width of $500 \; \rm ns$). Below we will be interested in
the parameter region of small kicking periods $T \lesssim 1$, and hence it is 
important to note that time is one of the best controlled 
experimental parameters, and 
kicking periods between about one hundred nanoseconds and a few hundred 
microseconds are available, with a maximal precision of a few 
tens of nanoseconds \cite{KRcontrol,OSR2003,WS2005}. For cesium atoms, 
this range corresponds to dimensionless kicking periods 
$T \approx 10^{-2} \ldots 20 $, and a maximal precision of 
$\delta T \gtrsim 10^{-3}$ \cite{WS2005}.  Atom optics 
experiments may be performed on two different atomic samples: laser 
cooled atoms and Bose-Einstein condensates. The  main difference 
is the initial width $\Delta p_0$ in momentum. For laser cooled atoms and 
in the best conditions, the initial width in momentum
corresponds to a few two-photon recoils units. For Bose-Einstein 
condensates  $\Delta p_0$ between 0.01 and 0.05 can be realized 
\cite{wilson,wilson1,nist,nist05}. Bose-Einstein condensates experience a 
nonlinear potential associated with the atom-atom interaction. However, 
letting the condensate expand a little before the 
actual kicking evolution allows one to reduce the atom-atom 
interactions to negligible values, with only slight changes 
in $\Delta p_0$ \cite{wilson}. The present analysis focuses on a sample
of laser cooled atoms with a large initial momentum distribution. In fact,  
this condition is more favorable for the realization of the  
ratchet discussed in this paper, because the sample explores a larger 
region of the classical phase space and therefore exploits the 
structure of phase space (a strange repeller, in the classical limit) 
induced by the evaporative cooling process.

The paper is organized as follows:
Section II analyzes the AOKR model and its evolution in phase space under
the double kicking perturbation. Section III  investigates different 
imperfections associated with the experimental realization. 
For instance a deep optical potential 
is required for laser-cooled atoms, and in such conditions spontaneous 
emissions become a non-negligible issue. In addition, 
fluctuations in the laser power and other sources of noise are 
included in the analysis. The final Section IV 
concludes with an outlook discussing the role of 
nonlinearity as present in experiments using a 
Bose-Einstein condensate.

%%%%%%%%%%%%%%%%%%%%%%%%%%%%%%%%%%%%%%%%%%%%%%%%%%%%%%%%%%%%%%%%%%%%%%%%%%%%%

\section{Model and properties}

In this section we introduce a kicked system that shows directed 
transport and in which the direction of the current can be controlled. 
This is done in a very simple way, we just have to duplicate 
the series of kicks in (\ref{eq:ham}) 
in a convenient fashion. This simplicity is 
essential for an efficient experimental implementation with cold atoms.

We consider a particle moving in one dimension 
[$x\in(-\infty,+\infty)$] in a periodically kicked potential.
The Hamiltonian reads
\[
{\cal H}_2(t)= \frac{p^2}{2}+ V_{\phi,\xi}(x,t),
\quad V_{\phi,\xi}=k \times
\]
\begin{equation}
\sum_{n=-\infty}^{+\infty} \left[ \delta(t-n T)
\cos(x)+ \delta(t-n T-\xi) \cos(x-\phi)\right],
\label{2kicks}
\end{equation}
where $T$ is the kicking period.
In fact, we propose an 
asymmetric kicking sequence. This is made out of two series of 
kicks with the same spatial and temporal periods, $2 \pi$ and 
$T=2\pi/\omega$, but shifted by a phase $\phi$
($0\leq \phi <2\pi)$ and a time $\xi$
($0\leq \xi <T$). 
Due to the spatial periodicity of the kicking potential $V_{\phi,\xi}$,
the one-cycle evolution (Floquet) operator 
\begin{equation}
\hat{\mathcal{U}}
=e^{-i (T-\xi) \hat{p}^2/2} 
e^{-ik\cos (\hat{x}-\phi)} 
e^{-i \xi \hat{p}^2/2} 
e^{-ik\cos (\hat{x})} 
\end{equation} 
induced by the Hamiltonian of (\ref{2kicks})  
commutes with spatial translations by multiples of $2\pi$.
As is well known from Bloch theory, this implies conservation of 
the quasi-momentum $\beta$, defined as the fractional part 
of the momentum $p$ ($0\leq \beta <1$) \cite{WGF2003}. 
For a given value of the quasi-momentum, the wave function of
the system is a Bloch wave, of the form $e^{i\beta x} \psi_\beta(x)$,
where $\psi_\beta(x)$ is a function of period $2\pi$. 
A generic wave function can then be written as a superposition of 
Bloch waves: $\psi(x)=\int_0^1 d\beta e^{i\beta x} \psi_\beta(x)$. 

Introducing the rescaled momentum variable $I=Tp$,
one can see that classical dynamics of model (\ref{2kicks}) depends on the
scaling parameter $K=kT$ (not on $k$ and $T$ separately).
The classical limit 
corresponds to $\hbar_{\rm eff}=T\to 0$,
while keeping $K=\hbar_{\rm eff} k$ constant. 

In order to simulate the evaporative cooling process in the quantum 
model we consider the projection over a subspace corresponding to 
the quantum levels that are below $p_c$ (in absolute value). 
In practice, this is implemented 
at each kick:
if we denote by $\Pop$ the projection operator on the 
interval $]-p_{c},p_{c}[$, the wave function after $n$ kicks is then given by
\begin{eqnarray}
\psi (p,n) = (\Pop \hat{\mathcal{U}}) ^{n} \psi(p,0) \;.
\label{floquet}
\end{eqnarray}
Note that quasi-momentum is still a conserved quantity.
In the classical case, we consider lost the particles that reach 
momentum $p$ such that $|p|>p_c$. 

We have checked in our numerical simulations 
that the dependence of the ratchet current on 
the cut-off value $p_c$ is weak, provided that $p_c\gg k$. 
Therefore, the ratchet current in this regime turns out to
be close to the current obtained in the Hamiltonian limit 
$p_c\to \infty$. On the other hand, the particle escape mechanism
strongly affects the phase space structure, leading, in the
classical limit, to the setting in of a strange repeller.

In the numerical simulations reported in this paper, we fix 
$K=7$, corresponding to the classically chaotic regime,   
$\xi=T/3$, and $p_c\hbar_{\rm eff}=15.2$. 
The initial state is given by a uniform 
mixture of the momentum states inside the interval 
$p\hbar_{\rm eff}\in [-1,1]$. Once the quasi-momentum is fixed, the number of 
momentum states in this interval is $\propto 1/\hbar_{\rm eff}$.
Moreover, we average numerical data over $10^3$ randomly 
chosen quasi-momenta. Classical averages are constructed from 
$10^7$ initial conditions randomly and uniformly distributed
inside the region $x\in [0,2\pi)$, $I=pT\in [-1,1]$. Note that with these
initial conditions and the above parameter values we are left
with approximately $35\%$ of the initial number of particles
at time $n=t/T=10$ and $10\%$ at $n=20$.  

\begin{figure}
\centerline{\epsfxsize=6.5cm\epsffile{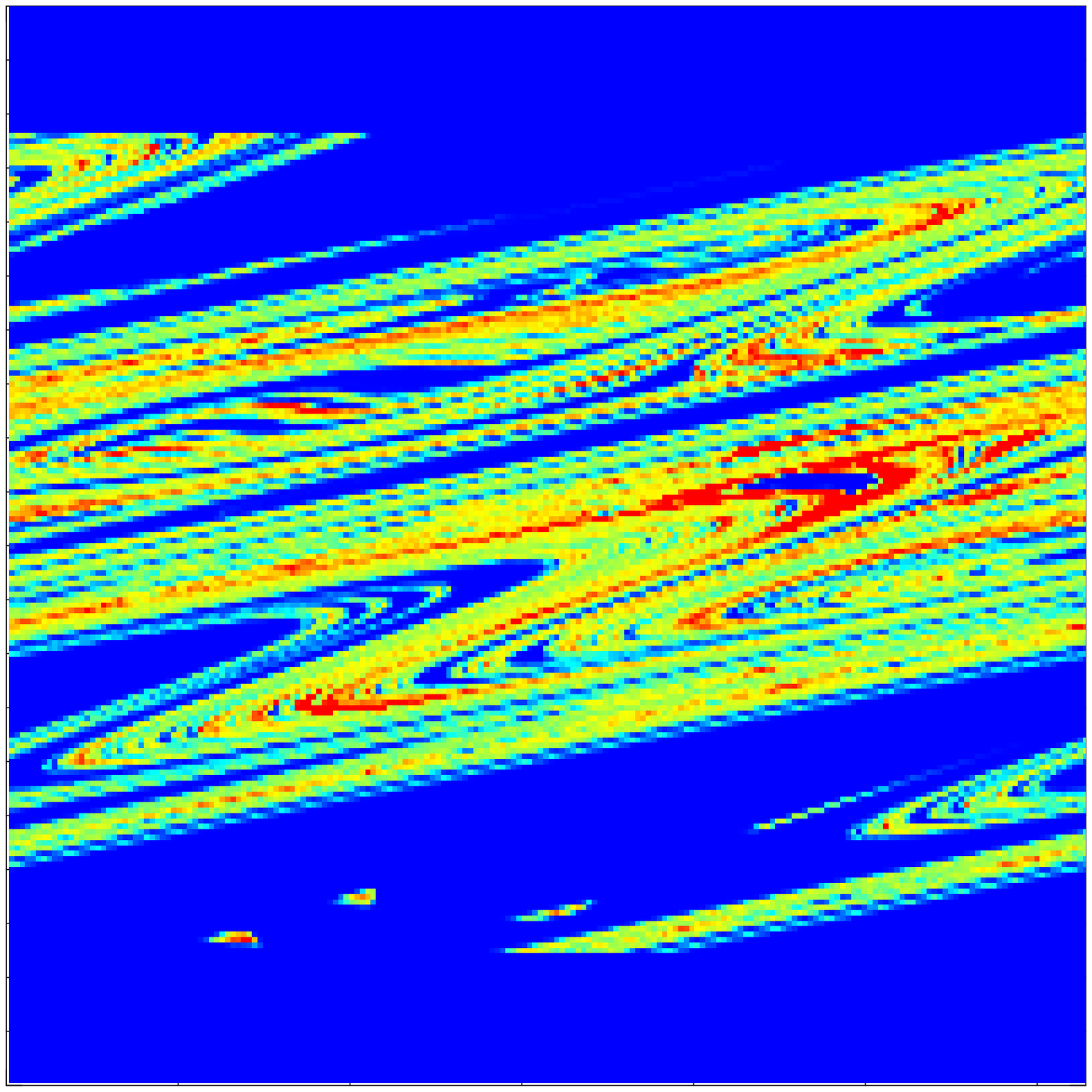}}
\centerline{\epsfxsize=6.5cm\epsffile{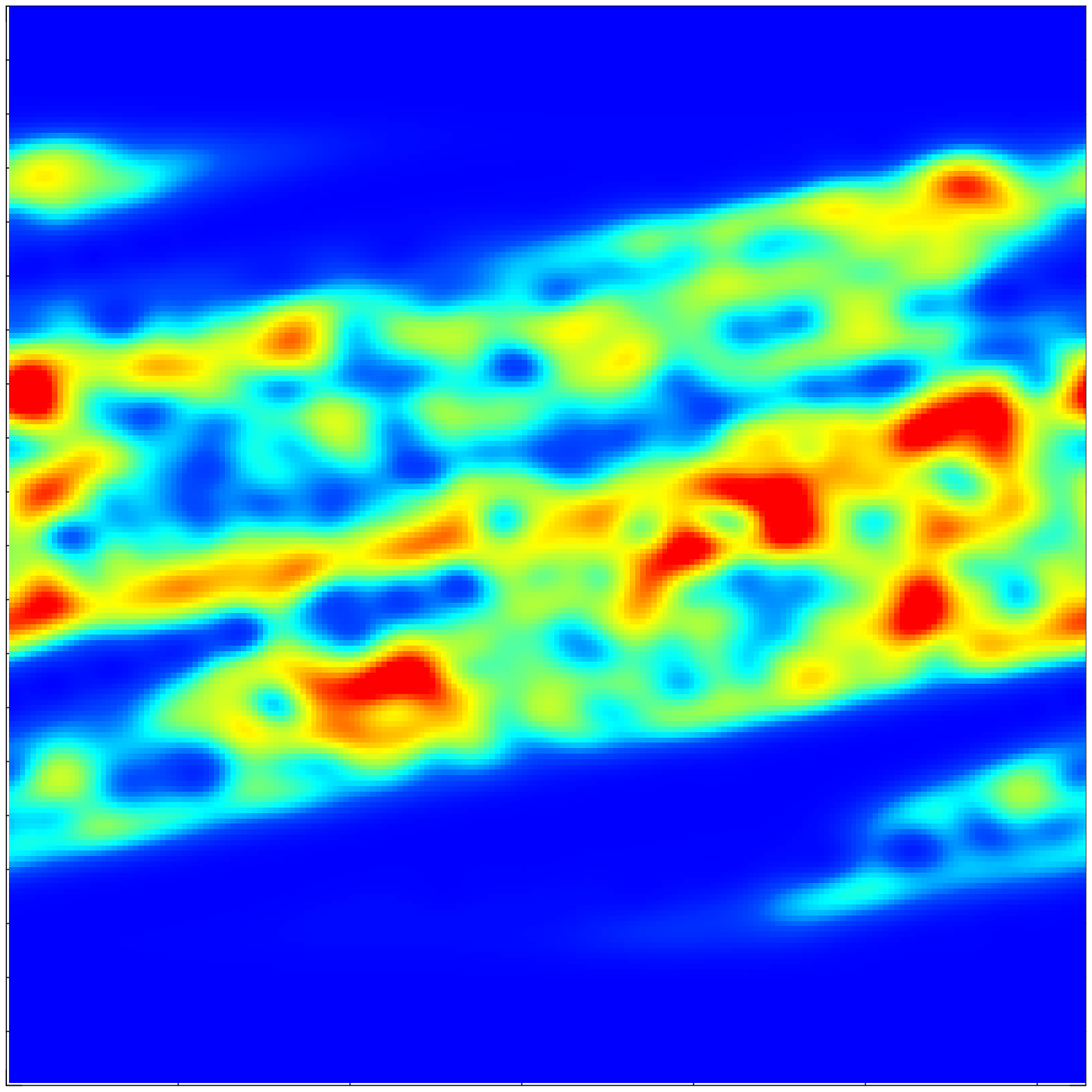}}
\centerline{\epsfxsize=6.5cm\epsffile{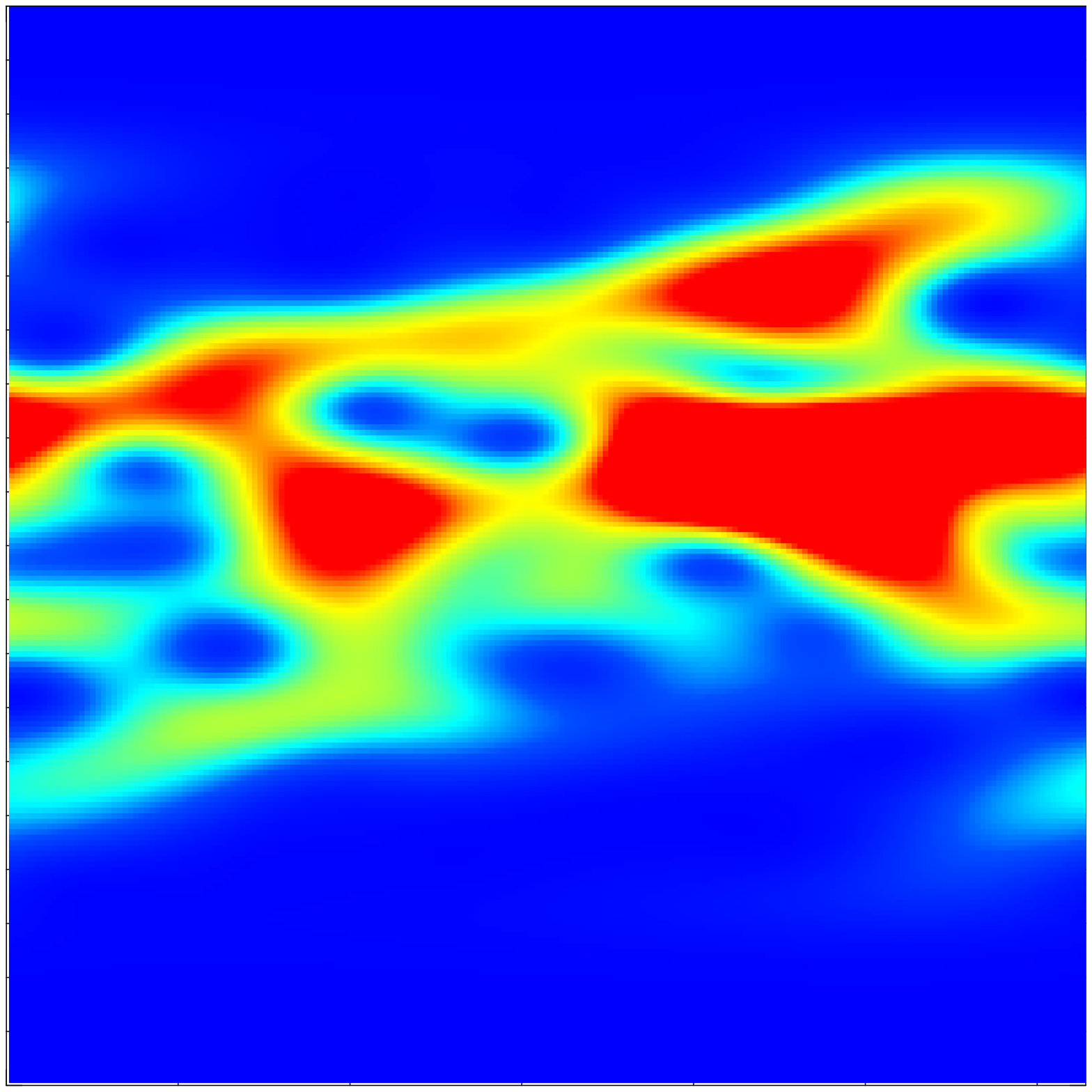}}
\caption{(Color online) Phase space pictures for 
$\phi=\pi/2$, at $n=20$: classical Poincar\'e sections (upper panel) 
and quantum Husimi functions at $\hbar_{\rm eff} \simeq 0.16$ 
(middle panel) and $\hbar_{\rm eff} \simeq 1$ (lower panel).
The displayed region is given by 
$I=pT \in [-20,20]$ (vertical axis) and $x \in [0,2 \pi)$ (horizontal axis).
Note that, to draw the attractor, $x$ is taken modulus $2\pi$.
The color is proportional to the density: blue for zero and 
red for maximal density.}
\label{portraits1}
\end{figure}

The appearance of a strange repeller in our model in
the classical limit is shown in the phase space portrait of 
Fig.~\ref{portraits1} (top), obtained for $\phi=\pi/2$ at $n=20$.
The three panels of Fig.~\ref{portraits1} 
correspond, from top to bottom,
to the classical Poincar\'e section
and the quantum Husimi function  at 
$\hbar_{\rm eff} \simeq 0.16$ and  
$\hbar_{\rm eff} \simeq 1$.
We can see a good agreement between the classical and the quantum 
phase space portraits. 
Quantum fluctuations smooth the fractal structure of the 
classical repeller on the scale of Planck's cell \cite{Graham}.
In the quantum case the values of $\hbar_{\rm eff}\simeq 0.16$ 
and $\hbar_{\rm eff}\simeq 1$ considered here (and suitable for a 
realistic experimental implementation) are not sufficiently
small to resolve the fractal structure at small scales.
However, the Husimi function shows clear similarities with the
underlying classical probability distribution.
Even for $\hbar_{\rm eff} = T \simeq 1$ 
the major features of the classical repeller (i.e., width in phase space 
and asymmetry) are visible. Parameter values and evolution time 
are suitable for the experimental measurement of the quantum 
probability distribution located on the underlying classical 
strange repeller. This is important because the appearance of 
strange sets (attractors or repellers) is a distinctive feature
of open chaotic systems.

\begin{figure}
\centerline{\epsfxsize=7cm\epsffile{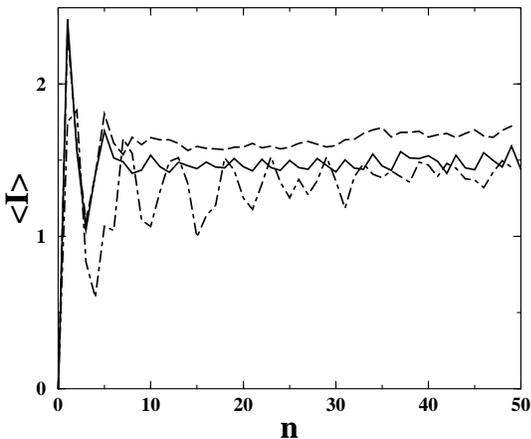}}
\caption{Average rescaled momentum $\langle I \rangle = \langle p \rangle T$ as a function
of the discrete time $n$, for the same parameter 
values as in Fig.~\ref{portraits1}.
The solid curve corresponds to the classical case, 
while the dashed curve corresponds to quantum results for 
$\hbar_{\rm eff} \simeq 0.16$ and the dot-dashed one to 
$\hbar_{\rm eff} \simeq 1$.}
\label{fig2}
\end{figure}

The repeller in Fig.~\ref{portraits1} is strongly asymmetric, 
suggesting directed transport, that is, $\langle p \rangle \ne 0$.
This is confirmed by the numerical data of Fig.~\ref{fig2}, 
where $\langle p \rangle$ is shown as a function of the time $n$.

We can explain the origin of the directed current present in our system 
by following the approach developed in \cite{Flach}. We have a 
classical time evolution given by, 
\begin{equation}
\ddot x + f_{\phi,\xi}(x,t) = 0,
\label{dyn}
\end{equation} 
where $f_{\phi,\xi}(x,t)= \partial V_{\phi,\xi}(x,t)/ \partial x$. 
To this equation we add a particle escape process consisting of cutting out 
the orbits that exceed a given value of the momentum $p=\dot x$. 
We are interested in symmetry transformations that leave 
Eq.~(\ref{dyn}) invariant but change the sign of $p$. In fact, 
if we assume that our system is chaotic  
we can generate for each orbit its $p$-reversed partner, which 
will explore the whole region embedding the 
chaotic trajectories. This amounts to saying that, 
being essentially equivalent, both orbits (and all of them) should 
have zero average momentum. If these symmetries are absent it is 
natural to conclude that a net $p$ (i.e., different from zero) 
can be generated. Thus, breaking all possible symmetries of this 
kind constitutes a good method to engineer ratchet systems. 
As the particle escape process introduced above is symmetrical with respect to 
$p$, we can neglect it in the following reasoning. 
%Nevertheless, 
%its introduction is relevant in order to analyze the stability of 
%our proposed ratchet. 
It is worth mentioning that all the symmetry 
considerations developed in this section translate almost immediately 
to the quantum case.

There are two general ways to change the sign of $p$:
\[
(I)\; \; \;
x \rightarrow -x+\alpha, \; \; \;
t \rightarrow t+\gamma,
\]
and 
\[
(II)\; \; \;
x \rightarrow x+\alpha, \; \; \;
t \rightarrow -t+\gamma.
\]
In order to leave Eq.~(\ref{dyn}) unchanged we need that 
$f_{\phi,\xi}(x,t)=-f_{\phi,\xi}(-x+\alpha,t+\gamma)$
holds for 
$(I)$, since $\ddot x \rightarrow - \ddot x$ under this 
transformation. If we apply twice transformation $(I)$ we obtain 
$f_{\phi,\xi}(x+\alpha/2,t) =f_{\phi,\xi}(x+\alpha/2,t+2\gamma)$. 
Since $f_{\phi,\xi}(x,t)$ is assumed to 
be bounded and periodic with zero mean, both in $x$ and $t$, 
$\gamma$ can only be an integer multiple of $T/2$ (including the 
$\gamma=0$ case). In turn, there are no restrictions on 
$\alpha$. 
On the other hand, for $(II)$ we need   
$f_{\phi,\xi}(x,t)
=+f_{\phi,\xi}(x+\alpha,-t+\gamma)$
(with a plus sign since now $\ddot x$ keeps its original sign). 
By applying twice transformation $(II)$ we obtain 
$f_{\phi,\xi}(x,t+\gamma/2) 
=f_{\phi,\xi}(x+2\alpha,t+\gamma/2)$. 
Following the same 
reasoning as before, $\alpha$ is fixed to integer multiples of 
$\pi$ (including $\alpha=0$) while there are no restrictions 
on $\gamma$.
Note that $(I)$ and $(II)$ are the only two symmetries that 
should be broken in order to find directed transport. Our choice of the 
potential (\ref{2kicks}) guarantees the possibility to break both of them.

In fact, we have that $f_{\phi,\xi}(x,t)=k \sum_{-\infty}^{+\infty}
[-\delta(t - nT) \sin(x) - \delta(t-nT-\xi) \sin(x-\phi)]$, 
and in the case of symmetry (I) we require that 
$f_{\phi,\xi}(x,t)=-f_{\phi,\xi}(-x+\alpha,t+\gamma)$.
We can take $\gamma=0$ without loss of generality since we only have 
a sum of delta functions in $t$, i.e. the sign change of 
$f$ induced by symmetry $(I)$ can only 
come from the first part of the transformation
($x\to -x+\alpha$).  
Therefore, we arrive at the conditions 
$\sin(-x+\alpha)=-\sin(x)$ and 
$\sin(-x+\alpha-\phi)=-\sin(x-\phi)$. 
These two conditions lead to 
$\alpha= l 2 \pi$ and $\alpha=l' 2 \pi + 2 \phi$,
with $l$ and $l'$ integers, 
and cannot be fulfilled together, except for 
$\phi=0$ or $\phi=\pi$. 
Therefore, symmetry $(I)$ is broken when $\phi\ne 0,\pi$. 

In the case of symmetry $(II)$, if we take 
$\alpha$ an odd multiple of $\pi$ then 
the sign of $f_{\phi,\xi}$ changes. 
Then, we are only left with $\alpha$
being an even multiple of $\pi$, i.e., we can take 
$\alpha=0$ without loss of generality. 
Moreover, we notice that if $\phi=0$ and 
$\alpha=0$ both kicks become the same in $x$ and therefore 
symmetry $(II)$ holds for any $\xi$, taking $\gamma=\xi$.
On the other hand, considering $\phi \ne 0$
we arrive at the conditions 
$\sum_{n=-\infty}^{+\infty} \delta(-t+\gamma-nT)=
\sum_{n=-\infty}^{+\infty} \delta(t-nT)$ and 
$\sum_{n=-\infty}^{+\infty} \delta(-t+\gamma-nT-\xi)=
\sum_{n=-\infty}^{+\infty} \delta(t-\xi)$, which 
imply $\gamma= l T$ and $\gamma=l' T +2 \xi$, with $l$ and $l'$ integers.
We conclude that, if $\phi\ne 0$, symmetry $(II)$ is broken 
when $\xi\ne 0, T/2$.  

In summary, both symmetries $(I)$ and $(II)$ are broken for 
$\phi\ne 0,\pi$ and $\xi\ne 0,T/2$. Hence two series 
of kicks are sufficient to observe the ratchet effect,
provided that these kicks are shifted both in 
space and in time, the shift in space being different from half
wave length and the shift in time being different from half
period.

\begin{figure}
\centerline{\epsfxsize=7cm\epsffile{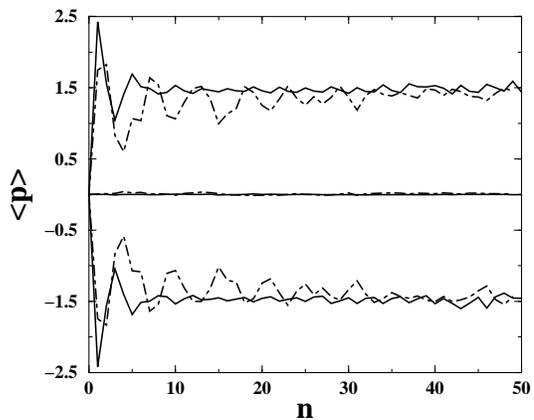}}
\caption{Average momentum $\langle p \rangle$ as a function
of $n$, for $\phi =\pi/2$ (positive values), 
$\phi=0$ (zero values), and $\phi=-\pi/2$ (negative values) 
Both the classical (solid curves) and the quantum 
case (dot-dashed curves, $\hbar_{\rm eff} \simeq 1$) are shown.
Note that at $\phi=0$ quantum and classical curves are
almost superimposed.}
\label{fig3}
\end{figure}

It is interesting to remark that current reversal can be engineered in a 
very simple way, by taking 
$\tilde \phi= - \phi$ instead of $\phi$ in (\ref{2kicks}).
Indeed, Eq.~(\ref{dyn}) is left unchanged when 
$x\to -x$, $t\to t$, and $\phi\to \tilde{\phi} = -\phi$,
while this transformation changes the sign of $p$.
We can see current inversion in Fig.~\ref{fig3}, both in 
the classical and in the quantum case, when 
$\phi=\pi/2\to \tilde{\phi}=-\pi/2$. Note that $\langle p \rangle =0$
at $\phi=0$, in agreement with the above symmetry considerations.

%%%%%%%%%%%%%%%%%%%%%%%%%%%%%%%%%%%%%%%%%%%%%%%%%%%%%%%%%%%%%%%%%%%%%%%%%%%%%

\section{Stability of the ratchet effect under imperfections}

\begin{figure}
\centerline{\epsfxsize=7cm\epsffile{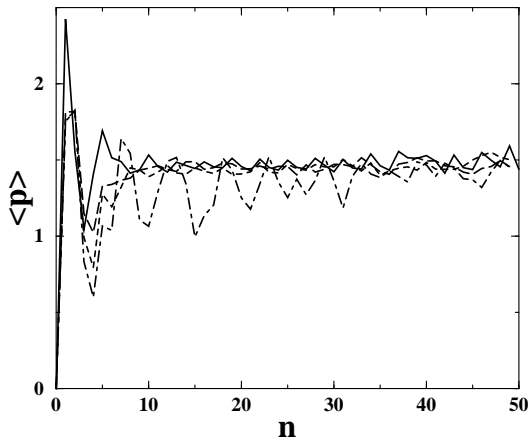}}
\caption{Average momentum $\langle p \rangle$ as a function
of the discrete time $n$, for the same parameters as in Fig.~\ref{fig2},
at $\hbar_{\rm eff}\simeq 1$. 
At each kick, the quasi-momentum can jump to any other possible value 
with probabilities 0 (dot-dashed curve), 0.2 
(dashed curve), and 0.5 (long dashed curve).
The solid curve corresponds to the classical case.} 
\label{fig4}
\end{figure}

The purpose of this section is to study the robustness of the ratchet 
effect introduced in this paper in the presence of typical sources 
of noise in cold-atom experiments. For the large
kicking strengths needed to guarantee clear signatures of a chaotic repeller,
spontaneous emission during the flashing of the optical lattice cannot
be ruled out \cite{KRexperiment}. 
Spontaneous emission can be effectively modeled by random jumps
in quasi-momentum \cite{WGF2003}. We test the influence of such random
changes in quasi-momentum on the results presented in the previous section.
That is to say, we repeat the previous calculations but letting at any kick 
the quasi-momentum randomly change with a probability of 0, 0.2 and 
0.5 (see Fig.~\ref{fig4}). In practice, it may jump to any 
possible value in the Brillouin zone with those probabilities. 
As can be seen, this additional randomness even helps to reduce 
fluctuations, and when the jump probability is different from zero there is 
a better convergence towards the classical result. 

We now investigate how different kind of errors affect the 
value of the ratchet current. More precisely, we compute 
the average current $\langle p \rangle_{\rm av}$, obtained after averaging 
$\langle p \rangle$ in the time interval $10\le n \le 20$, as a function
of the noise strengths associated to different noise sources.

\begin{figure}
\centerline{\epsfxsize=7cm\epsffile{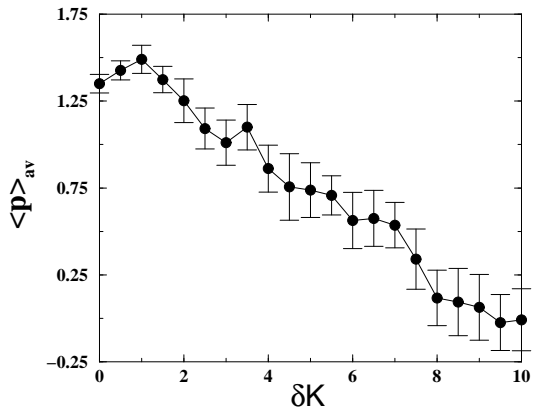}}
\caption{Average current $\langle p \rangle_{av}$ as a function   
of the noise $\delta K$ in the kick strength $K$, for parameter 
values as in Fig.~\ref{fig2}, at $\hbar_{\rm eff}\simeq 1$.}
\label{kick}
\end{figure}

First of all, we consider the effects of fluctuations
in the kicking strength. This is simulated 
by memoryless random errors of size $\delta K$ in the value of $K$:
the kicking strength $K_n$ at time $n$ is given by 
$K_n=K+(\delta K)_n$ where the noise value $(\delta K)_n$ is 
randomly drawn from a uniform distribution in the interval $[-\delta K, \delta K]$. 
It can be seen in Fig.~\ref{kick} that the ratchet effect is  
stable up to approximately 
$\delta K \simeq 2$, corresponding to a relative amplitude noise of
$\delta K /K \simeq 0.3$. 

\begin{figure}
\centerline{\epsfxsize=7cm\epsffile{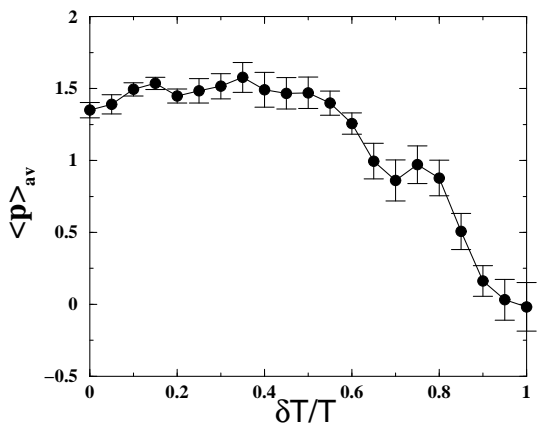}}
\caption{Average current $\langle p \rangle_{av}$ as a function   
of the relative error $\delta T/T$ in the kicking period $T$,
for parameter values as in Fig.~\ref{kick}.}
\label{tau}
\end{figure}

Since the ratchet mechanism described in the previous section
works the better the smaller we choose $\hbar_{\rm eff} = T$,
we consider possible fluctuations in the kicking period \cite{OSR2003} 
arising from the problem of controlling strong but narrow pulses 
in time with a high repetition rate. 
We model these imperfections as random and memoryless fluctuations 
in the period between consecutive kicks. This takes into account the fact that  
the timing of the kicks can suffer from uncontrollable 
variations.
As we can see from Fig.~\ref{tau}, stability is quite satisfactory  
when $\delta T/T \lesssim 0.5$, where $\delta T$ is the
size of the fluctuations and $T\simeq 1$.

\begin{figure}
\centerline{\epsfxsize=8cm\epsffile{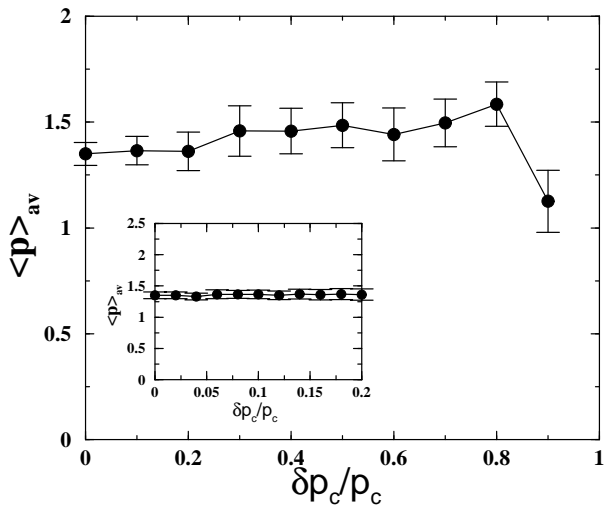}}
\caption{Average current $\langle p \rangle_{\rm av}$ as a function   
of the relative error $\delta p_c/p_c$ in the cut-off momentum $p_c$,
for parameter values as in Fig.~\ref{kick}.
A magnification of the figure for small values of $\delta p_c/p_c$
is shown in the inset.}
\label{cut}
\end{figure}

Finally, we consider the effect of an imprecision 
in the $p_c$ selection. 
This is modeled by random memoryless variations of the cut-off
value $(p_c)_n$ used at time $n$: $(p_c)_n=p_c+(\delta p_c)_n$,
with $(\delta p_c)_n\in [-\delta p_c,\delta p_c]$.  
Again the ratchet 
effect proves to be robust, as can be deduced from Fig.~\ref{cut}.
The results of this figure are in agreement with the previous
observation that the dependence of the ratchet current 
$\langle p \rangle$ on the cut-off value $p_c$ is weak
(under the condition $p_c\gg k$).

%%%%%%%%%%%%%%%%%%%%%%%%%%%%%%%%%%%%%%%%%%%%%%%%%%%%%%%%%%%%%%%%%%%

\section{Conclusions}

Considering a realistic experimental scenario, we
showed that a ratchet effect -- induced by a combination of a two-kick 
sequence as applied to  an open system -- is observable in an atom-optics
kicked rotor experiment. We also checked the robustness of the ratcheted
atomic  evolution under reasonable noise conditions. 

An interesting perspective would be to study the ratchet dynamics in 
a kicked Bose-Einstein condensate. Strong kicks may, however, lead to
thermal excitations out of equilibrium and destroy the condensate, rendering
the description by the usually applied Gross-Pitaevskii equation meaningless
\cite{zoller}.
We have verified that the ratchet evolution is preserved in the
presence of typical experimental nonlinearities. However,
a full treatment of a strongly kicked Bose-Einstein condensate 
remains a challenge for future work. 

%%%%%%%%%%%%%%%%%%%%%%%%%%%%%%%%%%%%%%%%%%%%%%%%%%%%%%%%%%%%%%%%%%%

We gratefully acknowledge support by the 
the MIUR COFIN-2004 and 2005, the EU Specific Targeted Research Project OLAQUI,
and the Alexander von Humboldt Foundation (Feodor-Lynen Program).
G.G.C. greatfully acknowledges support by Conicet
(Argentina).

%%%%%%%%%%%%%%%%%%%%%%%%%%%%%%%%%%%%%%%%%%%%%%%%%%%%%%%%%%%%%%%%%%%

\end{document}